\documentclass[english]{iopart}
\usepackage[T1]{fontenc}
\usepackage[latin1]{inputenc}
\usepackage{graphicx}
\usepackage{amssymb}

\makeatletter

\@ifundefined{textcolor}{}
{%
 \definecolor{BLACK}{gray}{0}
 \definecolor{WHITE}{gray}{1}
 \definecolor{RED}{rgb}{1,0,0}
 \definecolor{GREEN}{rgb}{0,1,0}
 \definecolor{BLUE}{rgb}{0,0,1}
 \definecolor{CYAN}{cmyk}{1,0,0,0}
 \definecolor{MAGENTA}{cmyk}{0,1,0,0}
 \definecolor{YELLOW}{cmyk}{0,0,1,0}
 }

\usepackage{subfigure}\usepackage{epsfig}\usepackage{dcolumn}
\usepackage{bm}\usepackage{amsfonts}\usepackage{babel}

\newcommand{\bra}[1]{\langle #1 |}
\newcommand{\ket}[1]{| #1 \rangle}
\newcommand{\be}{\begin{equation}}
\newcommand{\ee}{\end{equation}}
\newcommand{\beq}{\begin{eqnarray}}
\newcommand{\eeq}{\end{eqnarray}}
\makeatother

\makeatother

\usepackage{babel}

\begin{document}

\title[FFLO oscillations and magnetic domains in the bond-charge Hubbard model]{Fulde Ferrell Larkin Ovchinnikov oscillations and magnetic domains
in the Hubbard model with off-diagonal Coulomb repulsion}

\author{A Anfossi$^1$, C Degli Esposti Boschi$^2$\footnote{On leave from CNISM, Unità di Ricerca del Dipartimento di Fisica dell'Università di Bologna, where this work was started.} and A Montorsi$^1$}

\address{$^1$ Dipartimento di Fisica del Politecnico, corso Duca degli Abruzzi
24, I-10129 Torino, Italy}
\address{$^2$ CNR-IMM, Sezione di Bologna, via Gobetti 101, I-40129, Bologna, Italy}
\eads{\mailto{alberto.anfossi@polito.it}, \mailto{degliesposti@bo.imm.cnr.it}, \mailto{arianna.montorsi@polito.it}}

\date{\today}
\begin{abstract}
We observe the effect of non-zero magnetization $m$ onto the superconducting
ground state of the one dimensional Hubbard model with off-diagonal
Coulomb repulsion $X$. For $t/2\lesssim X\lesssim2t/3$, the system
first manifests Fulde-Ferrell-Larkin-Ovchinnikov oscillations in the
pair-pair correlations. For $m=m_{1}$ a kinetic energy driven macroscopic
phase separation into low-density superconducting domains and high-density
polarized walls takes place. For $m>m_{2}$ the domains fully localize,
and the system eventually becomes a ferrimagnetic insulator. 
\end{abstract}

\maketitle

\section{Introduction}

\label{secI}

In the century-long search for a mechanism which unveils the origin
of superconductivity in so many different materials, a key ingredient
is to understand the role of magnetic correlations in the formation
of the superconducting (SC) pairs. Apart form the Meissner effect,
different magnetic effects have been described in superconductors,
such as the ferromagnetic to superconductor transition in heavy fermions
compounds\cite{SAXal}, or the magnetic field driven SC-insulator
transitions in some two-dimensional high-$T_{c}$ samples\cite{MCG}.
The very recent discovery of high-$T_{c}$ superconductivity in iron
based layered pnictides \cite{KWHH}, and the observation of coexisting
micro- and/or mesoscopic SC and magnetically ordered domains\cite{PARKal,LANGal}
which are reminiscent of the stripes characteristic of cuprates, confirms
the suggestion that superconducting and magnetic order are intimately
related, and can be crucial for each other's stability \cite{MONal}.
On the theoretical side, it has also been predicted that SC correlations
in presence of non-zero magnetization can rearrange their spatial
modulation exhibiting Fulde-Ferrell-Larkin-Ovchinnikov(FFLO) oscillations
\cite{FFLO}, which are expected in case of strongly anisotropic (for
instance one-dimensional) systems. It is a remarkable very recent
achievement their observation in both heavy fermions systems and one
dimensional Fermi gases\cite{LIal,KOUal}.

It is generally accepted that the above behaviours are to be ascribed
to the presence of electronic correlations in the different systems.
The reference model to deal with electron-electron interaction on
lattices is given by the Hubbard Hamiltonian, in which the main contribution
of the Coulomb repulsion to the model Hamiltonian is identified with
the on-site interaction of electrons with opposite spins ($U$ term).
The presence of a SC phase at $U>0$ for this model is still matter
of debate, though some encouraging results have been achieved (see
for instance \cite{EIBA} and the discussion therein). Many extensions
of the Hubbard model have been proposed, based on the inclusion of
the first contributions to the Coulomb repulsion other than the $U$
term. Among them, a generalization motivated independently by Hirsch\cite{HIR}
and Gammel and Campbell \cite{GACA} amounts to include the nearest
neghbors off-diagonal interaction $X$, which turns out to modulate
the hopping term depending on the occupations of the two sites. In
this case, the model Hamiltonian reads 
\begin{equation}
H = -\sum_{<i,j>\sigma}[1-x\:(n_{i\bar{\sigma}}+n_{j\bar{\sigma}})]c_{i\sigma}^{\dagger}c_{j\sigma}+u\sum_{i}n_{i\uparrow}n_{i\downarrow}
   +h\sum_{i}(n_{i\uparrow}-n_{i\downarrow})\;\label{eq:1}
\end{equation}
where $c_{i\sigma}^{\dagger}$ creates a fermion with spin $\sigma=\{\uparrow,\downarrow\}$,
$\bar{\sigma}$ denoting the opposite of $\sigma$, $n_{i\sigma}=c_{i\sigma}^{\dagger}c_{i\sigma}$
is the $\sigma$-electron charge and $\langle ij\rangle$ stands for
nearest-neighboring sites. The parameters $u$ and $x$ are the dominant
diagonal and off-diagonal contribution coming from Coulomb repulsion,
and the lower case symbols denote that these coefficients have been
normalized in units of the hopping amplitude. Here we have also included
an external magnetic field $h$. Moreover, $N$ is the number of electrons
on the $d$-dimensional $L$-sites lattice, so that $n=N/L$ is the
average filling value per site. The model has been extensively studied
in the literature at $h=0$ \cite{HBC}. In particular, since $H$
is not invariant under particle-hole transform, it has been proposed
to model a theory of hole superconductivity.\cite{HIR}\\
 While the effect of the off-diagonal Coulomb repulsion (also known
as bond-charge interaction) has long be disregarded since it is usually
smaller than the diagonal contributions of Coulomb interaction (Hubbard
and extended Hubbard models), it has by now become clear that what
matter is instead its amplitude compared with that of the hopping
term. In fact already for $x\gtrsim1/2$ a scenario quite different
from that of the Hubbard limit emerges \cite{ADMO,AAA} even in $d=1$.
In particular, for not too large on-site Coulomb repulsion $u<u_{c}(x)$
a new metallic phase appears, characterized by dominant SC correlations
and nanoscale phase separation\cite{ADM} (NPS). The latter amounts
to the coexistence of two conducting phases with different densities\cite{MON},
and is driven by the charge degrees of freedom. The actual size of
the coexisting domains turns out to be microscopic at $h=0$, and
to increase in relation to the imbalance in spin orientation\cite{ABM},
suggesting an interplay of charge and spin degrees of freedom in the
ground state. Since none of the two coexisting phases is separately
SC, it is reasonable to interpret this emergent SC behaviour as a consequence
of the microscopic domain size induced, in turn, by spin rearrangement;
we are in presence of magnetic mediated superconductivity. A further
confirmation of this hypothesis comes from a very recent result \cite{RDM}
which shows that the critical line for the SC transition can be recovered
exactly (see figure \ref{fig1}, left panel) if assuming short range
antiferromagnetic correlations between just the single electrons.
It must be noticed that the appearance of antiferromagnetic and metallic
behaviour at half-filling, in a regime of moderate interaction, as
well as its interplay with SC properties, is somehow reminiscent of the
physics characterizing iron pnictides in higher dimension. 

To better understand the role of magnetic correlations in ground-state
of Hamiltonian (\ref{eq:1}) and their interplay with the charge degrees
of freedom, here we shall investigate --through a detailed density-matrix
renormalization group (DMRG) numerical study \cite{DMRG}-- the consequences
of the inclusion of the magnetic field as to the onset of the SC phase.
In principle, increasing the magnetic field should force the spin
degrees of freedom to align to the external field, so that the underlying
phase separation (PS) in the charge degrees of freedom should emerge
at a macroscopic level, wiping out the SC properties. In fact, what
we will see is that the SC pairs can survive even at non-zero magnetization,
acquiring a modulation of FFLO type \cite{FFLO}. For higher values
of the external field however, the density domain texture of the ground
state emerges, and the SC pairs become first confined to the low density
domains only, while the high density domains behave as polarized walls.
Finally, above an upper value of the magnetic field superconductivity
disappears and the ferromagnetic domains localize: the system becomes
a ferrimagnetic insulator.

In section II we describe some features of the model Hamiltonian,
investigating in particular the role of the kinetic energy as to the
onset of the different transitions. In section III we investigate
the static spin and charge structure factor dependence on the magnetization.
The study of the Luttinger charge ($K_{\rho}$) and spin ($K_{\sigma}$)
parameters allows to identify the presence of superconducting metallic
and insulating phase. In section IV we explore the spatial dependence
of density and magnetization with increasing the magnetic field. In
section V we study the dependence of various type of pair-pair correlations
and give evidence of the related emergence of FFLO oscillations. Finally
in section VI we discuss the results and give some conclusions.

\section{The bond-charge Hubbard model }

\label{secII}

\subsection{Phase diagram at $h=0$ }

\begin{figure}
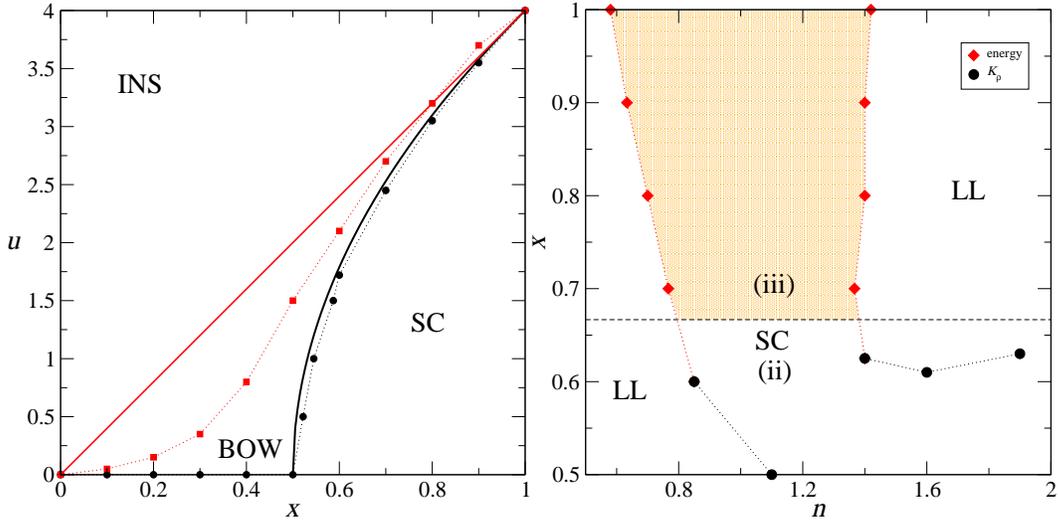

\includegraphics[width=7cm,clip]{fig1l}\includegraphics[width=7cm,clip]{fig1r}
\caption{Ground state phase diagram of $H$ in one dimension (in the absence of external magnetic field). INS denotes the insulating phase,
LL denotes the phase described by a Luttinger Liquid while BOW denotes the phase supporting bond order (charge density) wave \cite{GIA}. Symbols
(with joining dots) denote numerical results. Left panel: half-filling;
continuous lines denote the analytical expressions for the critical
lines obtained in \cite{RDM}. Right panel: $u=1$ (the legend refers
to the methods used in \cite{ADM} to locate the transition points).
Dashed line marks $x=2/3$.\label{fig1}}
\end{figure}

The ground-state phase diagram of Hamiltonian (\ref{eq:1}) at $h=0$,
as obtained numerically in \cite{ADMO,AAA,ADM,RDM} is reported
in figure \ref{fig1}. On the left-hand side this is given at fixed
filling ($n=1$) and varying $u>0$, $x$, whereas on the right-hand
side it is given at $u=1$, varying $n$ and $x$. Various phases
can be recognized: in particular a Luther-Emery liquid phase \cite{GIA} with
dominant superconducting correlations (SC) takes place in a range
of filling values which depends on $x$. Within such phase,
one can distinguish three different regimes: 
\begin{itemize}
\item {(i)}For $0<x\lesssim1/2$ the transition to the SC phase possibly
takes place for $n>1$ and not too large $u$ as soon as $x\neq0$,
in agreement with bosonization predictions \cite{JAKA} (not shown).
\item {(ii)} For $1/2\lesssim x\lesssim2/3$ we are in an intermediate
regime. The SC phase still appears for $0<u\leq u_{c}(x,n)$ in a
wider range $n_{l}\leq n\leq2$ with $n_{l}\leq1$. 
\item {(iii)}For $2/3\lesssim x\leq1$ nanoscale PS (NPS) is observed
in the SC phase as a texture of two fluids of different densities
$n_{l}$ and $n_{h}$, which coexist up to $u_{c}(x,n)$ for a range
of filling values $n_{l}\leq n\leq n_{h}<2$, with $n_{l}\leq1$. 
\end{itemize}

\subsection{Role of kinetic energy}

\begin{figure}
\includegraphics[width=7cm]{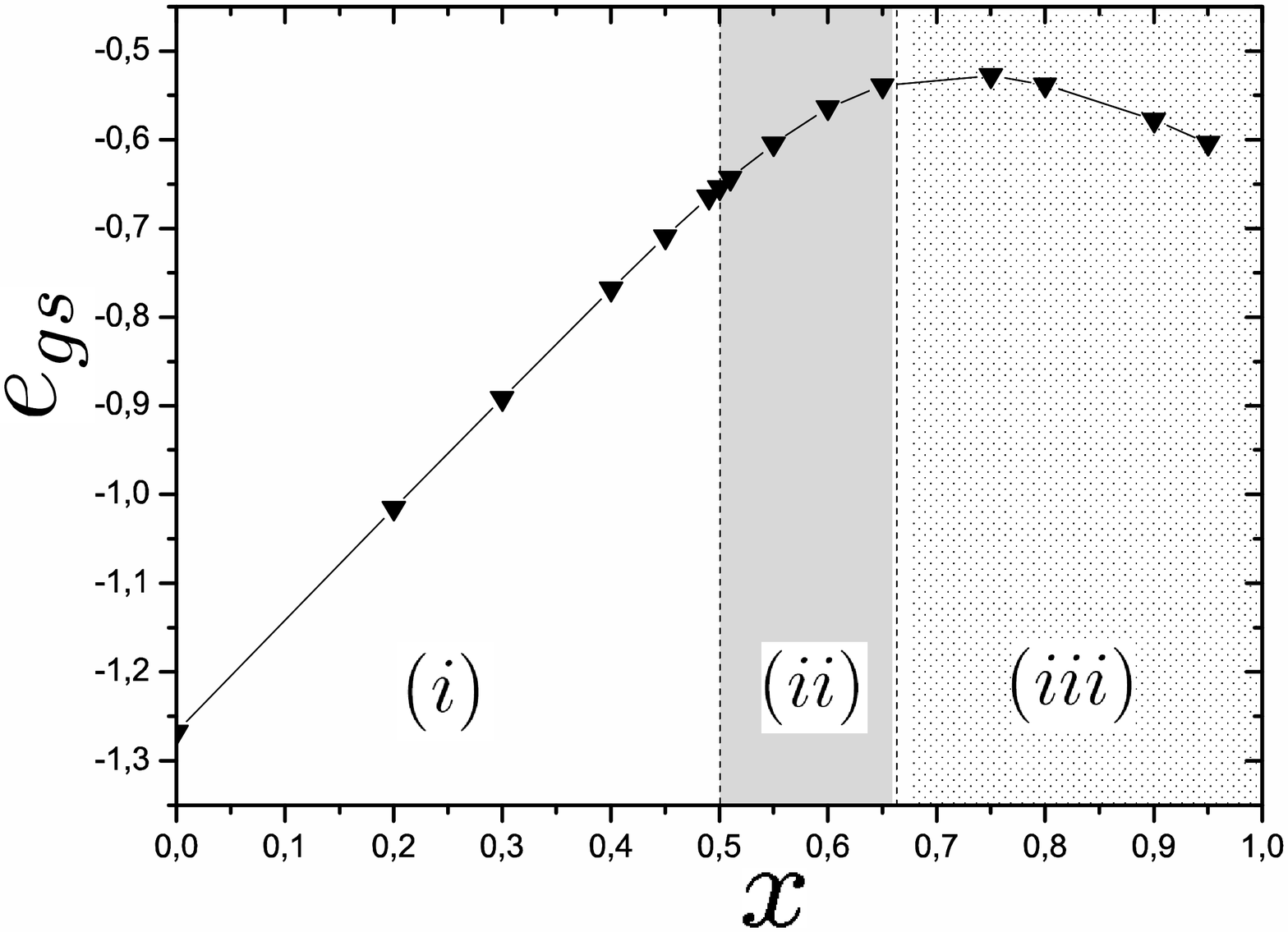}\includegraphics[width=7cm]{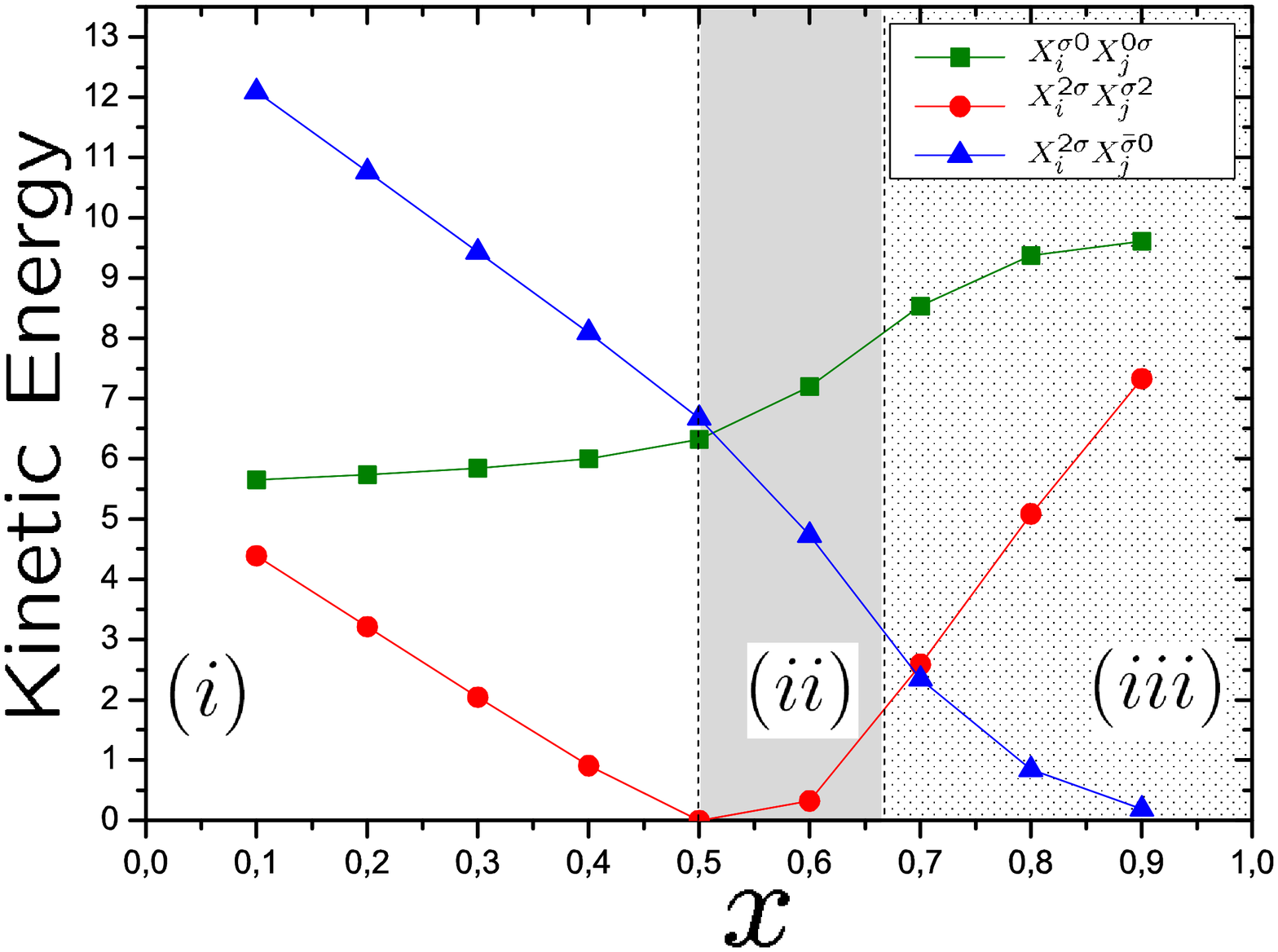}
\caption{The case of vanishing magnetization at half-filling,
$u=0$. Left: Ground-state energy vs $x$. Right: the three contribution
to kinetic energy in (\ref{eq:HAM_SA}) plotted vs $x$. In both
cases the three different regions (with white, gray, and dotted background)
correspond to the three different behaviour described in section \ref{secI}:
Hubbard-like, SC, NPS.\label{fig2}}
\end{figure}

The crossover among the above different regimes can be observed for
instance on the ground-state energy $e_{gs}$, plotted in figure \ref{fig2}(left)
at half-filling and $u=0$ versus $x$. With increasing $x$, $e_{gs}$
turns out to increase linearly from the value $e_{k}(x=0)=-4/\pi$
(non interacting electrons with spin) to the value $e_{k}\approx-2/\pi$
reached for $x=1/2$, in agreement with bosonization predictions.
Being the latter coincident with the energy of a system of spinless
fermions at quarter filling, from this point on the system becomes
unstable with respect to PS into two regions at different densities
(approximately $1/4$ and $3/4$) in region (ii), and correspondingly
$e_{gs}$ is seen to deviate from the linear increasing, reaching
its maximum for $x\approx2/3$. Above such value, the system enters
region (iii), and the ground-state energy begins to decrease, to gain
the value $-2/\pi$ precisely at $x=1$. At this point the system
can be regarded either as describing a fluid of spinless fermions
moving in a background of empty and doubly occupied sites\cite{AAS},
or as two coexisting fluids of single electrons moving in a background
of empty and doubly occupied sites \cite{MON}, in which case $n_{l}=0.5$
and $n_{h}=1.5$.

The significance of the points $x=1/2$ and $x=2/3$ at $u=0$ becomes
evident when representing the Hamiltonian in terms of on-site Hubbard
projectors, which are operators defined as $X_{i}^{\alpha\beta}\doteq\ket{\alpha}_{i}\bra{\beta}_{i}$.
Here $\ket{\alpha}_{i}$ are the states allowed at a given site $i$,
and $\alpha=\{0,\uparrow,\downarrow,2\}$ {[}$\ket{2}\equiv\ket{\uparrow\downarrow}${]}.
In this case the nonvanishing entries of the Hamiltonian matrix representation
are read directly as the nonvanishing coefficients of the projection
operators. When rewritten in terms of these operators, $H$ turns
out to be a subcase of the more general Hamiltonian introduced by
Simon and Aligia,\cite{SIAL}  

\begin{equation}
H_{BC}=-\sum_{<ij>\sigma}\left[ X_{i}^{\sigma0}X_{j}^{0\sigma}+t_{x}X_{i}^{2\sigma}X_{j}^{\sigma2}
+s_{x}(X_{i}^{\sigma0}X_{j}^{\bar{\sigma}2}-X_{i}^{2\sigma}X_{j}^{0\bar{\sigma}})\right]+u\sum_{i}X_{i}^{22}
\label{eq:HAM_SA}\end{equation}
in which $t_{x}=1-2x$ and $s_{x}=1-x$. Besides $u$, the behaviour
of $H_{BC}$ is determined by the strength of $t_{x}$ and $s_{x}$.
In particular, $x=1/2$ implies $t_{x}=0$, whereas below and above
such value $t_{x}$ changes sign. A negative $t_{x}$ induces frustration
in the motion of pairs, since it favors the presence in the ground state of momenta
close to zero. A positive $s_{x}$ term instead drives
the pairs hopping favoring the modes with momenta close to the
edge of the Brillouin zone. Hence we expect that for $|t_{x}|\gtrsim|s_{x}|$
--in our case $x\gtrsim2/3$-- the mobility of the pairs becomes favored
in the system for $u\leq u_{c}(x)$. This is summarized in the right-hand
side of figure \ref{fig2}, where the three contributions to the kinetic
energy in (\ref{eq:HAM_SA}) are plotted separately at $u=0$. It
is seen that in the three regions the relative weights of these contributions
to the total kinetic energy are ordered in different ways. In particular,
while at $x\leq1/2$ the term with coefficient $s_{x}$, which does
not conserve the number of pairs, plays the dominant role, for $x\geq1/2$
this role is taken by the term describing the mobility of single electrons
in a background of empty sites (coefficient $1$), conserving the
number of doublons; for $x\gtrsim2/3$ the term with coefficient $s_{x}$
becomes the smallest, to vanish exactly at $x=1$. In practice, a
consequence of the above behaviour is that for $x\geq2/3$ in the ground
state the number of pairs is conserved, and the fluid behaves as a
system of $N_{s}$ single electrons moving in a background of $L-N_{s}$
empty and doubly occupied sites. This is shown for instance in the
charge structure factor $N(q)=\sum_{i}(\langle n_{i}n_{i+r}\rangle-\langle n_{i}\rangle\langle n_{i+r}\rangle){\rm{e}}^{{\rm{i}} q r}$,
which correspondingly manifests a feature at $q=2\pi N_{s}/L$ 

The reasons for the further choice of NPS, with respect to macroscopic
PS, are to be found in the behaviour of the spin degrees of freedom,
and their interplay with charge degrees of freedom. Indeed the term
with coefficient $s_{x}$ tends to favor antiferromagnetism at short
range in an appropriate background of empty and doubly occupied sites.
It is the surprising result of a very recent paper \cite{RDM} that
this simple assumption allows to map the Hamitonian in the SC regime
into an effective XY model in a transverse field, recovering the critical
line shown in figure \ref{fig1} as the factorization line of the XY
model. Remarkably, such line is not critical in the XY model: it is
just the coupling of it to the charge degrees of in the original model
which determines a change of phase.

\subsection{Results at $h\neq0$}

\label{secIIB} 
\begin{figure}
\includegraphics[width=9cm]{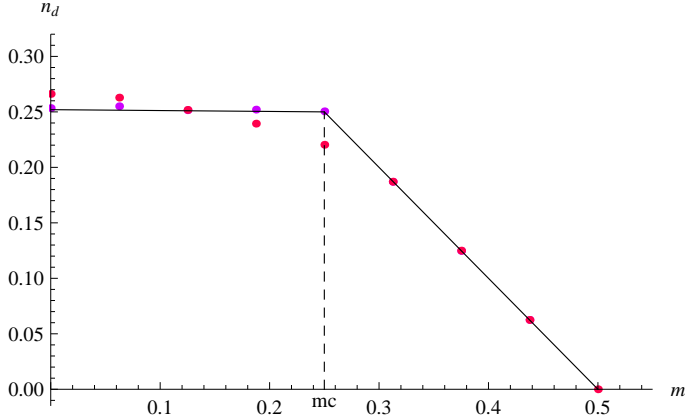}
\caption{Number of doubly occupied sites versus magnetization at two different
values of $x$, inside region (ii) (pink dots) and (iii) (violet dots) respectively. Continuous
line is the theoretical prediction at $x=0.8$ ($s_{x}=0$); dashed
line is the critical magnetization at which the breached pair regime
is entered in that case.\label{fig3}}
\end{figure}

At $h\neq0$ there are results available both at $x=1$\cite{DOMO},
which case can be treated exactly also at non-zero temperature, and
at $x\neq1$\cite{KESC}, where the numerical method employed required
the presence of a non-vanishing temperature. A low temperature peak
in the specific heat, to be ascribed to the excitation of the spin
degrees of freedom, is observed for $x\leq1/2$ already at vanishing
field, and at non-vanishing field for $x\lesssim1$.\\
Also, numerical results at $T=0$ have been obtained more recently
for imbalanced species of ultracold fermionic atoms, both in case
of attractive \cite{WADU}, and in case of repulsive \cite{ABM} $u$.
In this latter case it was realized that a non-trivial consequence
of the phase coexistence in region (iii) is that within the NPS phase
the number of pairs $n_{d}$ is constant with increasing $m$, up
to a critical magnetization $m_{c}=n-n_{\downarrow}=n-2n_{d}$ at
which the doublons begin to break, entering a regime of breached pairs.
This is shown by the numerical data reported in figure \ref{fig3},
where also the analytical value obtained at the same $t_{x}$ values
assuming $s_{x}=0$ are reported. It is seen that while in region
(iii) the numerical data at $s_{x}=0$ coincide with the theoretical
curve already at $m=0$, in the intermediate region (ii) this happens
only at large enough values of $m$ ($m>m_{c}$). Also the size of
the coexisting domains, microscopic for $h=0$, is observed to become
macroscopic at appropriate non-zero magnetization in both regions. 

\subsection{DMRG simulations}

Some observables considered in this work are better studied with periodic
boundary conditions, while others are obtained numerically with more
precision employing open boundary conditions. In both cases we retained
up to 768 optimized DMRG states, performing usually three finite-system
sweeps to enforce ground-state convergence. With more than 100 sites,
the truncation errors in the density matrix weight remain $O(10^{-6})$
or smaller. As far as the dependence on the system size is concerned,
due to the large number of points selected in the phase diagram we
decided to report the results for a fixed chain length L=160, unless
otherwise specified. However, for some selected points we carried
out a preliminar analysis of system size dependence (not reported
here) and observed that the behaviour of the various correlation functions
was essentially the same at different values of (sufficiently large)
L. Moreover, due to the onset of PS for some parameters value (see
below), it could be difficult to reproduce the same qualitative spatial
pattern by varying the system size. Hence the data presented at fixed
length are chosen to be representative of the physical behaviour seen
also at smaller sizes.

\section{Luttinger exponents: charge and spin structure factor at $h\neq0$}

We have seen that at non-zero magnetization the structure of the ground
state in region (iii) does not change --for what concerns the presence
of empty and doubly occupied sites-- up to the value $m_{c}$. To
some extent, this observation holds even inside region (ii). This
is consistent with the possibility that an external magnetic field
simply changes the polarization of the single electrons in the ground
state. At non zero magnetization however the electrons should arrange
differently regarding their pairing property. On general grounds, one
of the two following scenarios is expected to hold above a first critical
field $h_{c_{0}}$ at which the system begins to magnetize: either
the SC properties are lost, or the SC pairs acquire a non-vanishing
momentum and FFLO oscillations are observed \cite{FFLO}, due to the
presence of polarized single electrons which do not pair. 

\begin{figure}
\includegraphics[width=7cm]{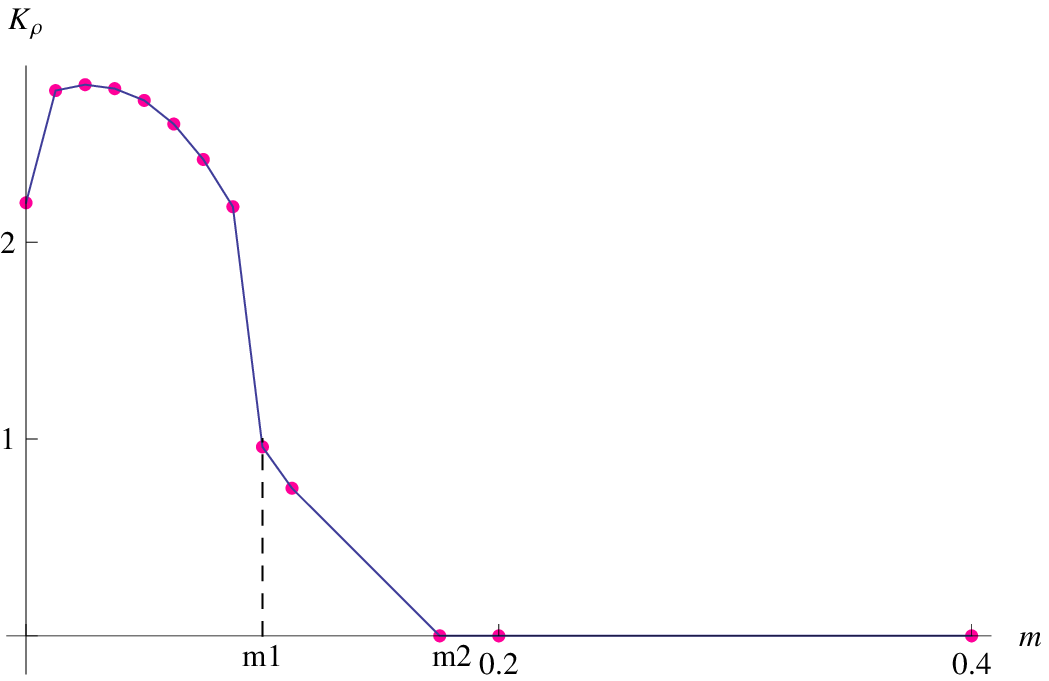}\includegraphics[width=7cm]{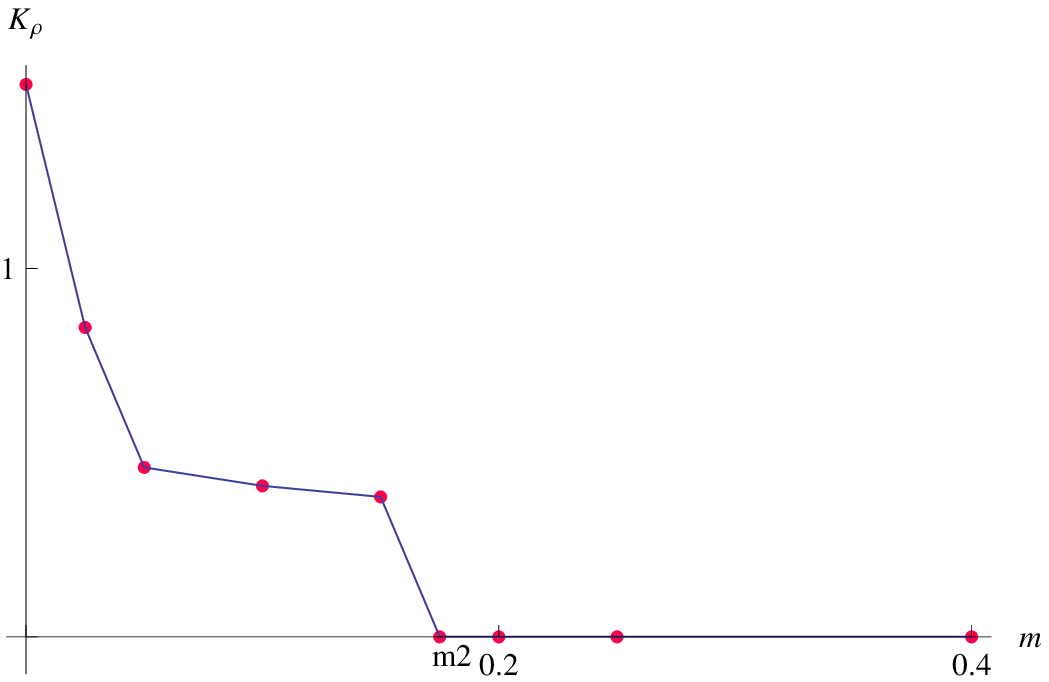}
\caption{Luttinger exponent (see text) $K_{\rho}$ at $u=0$,
$x=0.6$ (region (ii), left), and $x=0.8$ (region (iii), right) as
obtained from DMRG simulations (L=160). The two discontinuities identify
the critical values $m_{1}$ and $m_{2}$ discussed in the text.\label{fig4}}

\includegraphics[width=9cm]{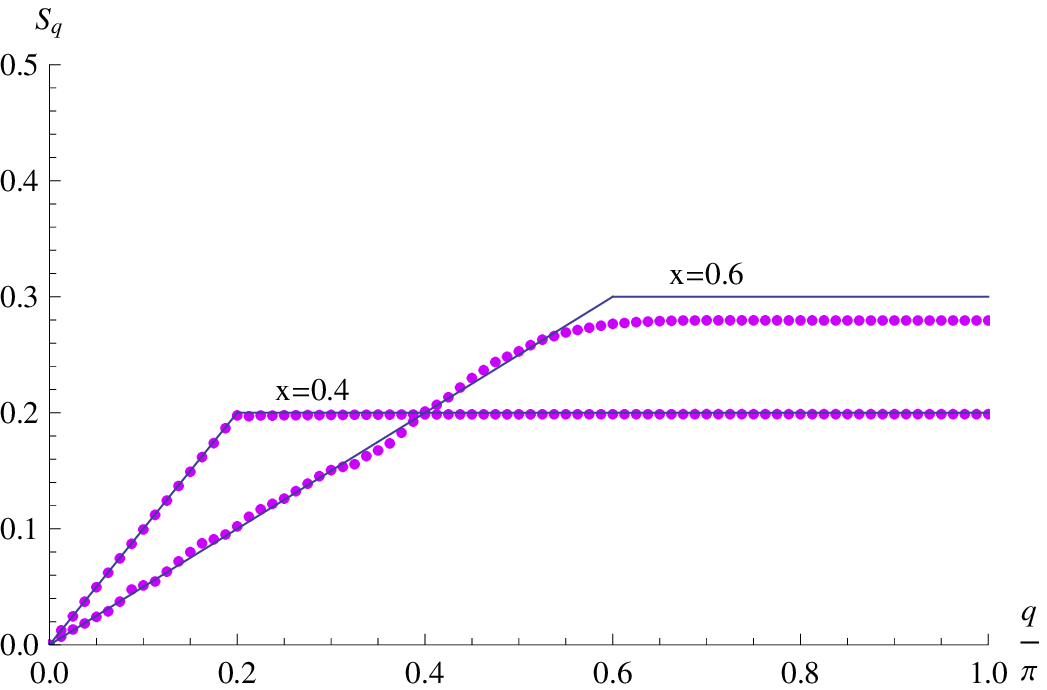}
\caption{Spin structure factor at high magnetization $m>m_{c}$
($m=0.4$), for $x=0.4$ and $x=0.6$ as obtained from DMRG simulations
($L=160)$. The crossover from a spinful to a spinless regime is shown:
continuous lines represent the theoretical prediction for a spinful
fluid at half filling ($x=0.4$); and a spinless fluid of $2m$ fermions
($x=0.6$).\label{fig5}}
\end{figure}

In order to explore what happens to the SC properties of the ground
state in the two regions (ii) and (iii) at non-zero magnetization,
we first evaluate the static charge $N(q)$ and spin $S(q)$ 
($S(q)=\sum_{i}(\langle s_{i}^{(z)}s_{i+r}^{(z)}\rangle-\langle s_{i}^{(z)}\rangle\langle s_{i+r}^{(z)}\rangle){\rm{e}}^{{\rm{i}} q r}$,
with $s_{i}^{(z)}\doteq(n_{i\uparrow}-n_{i\downarrow})$) structure
factors at different values of the average magnetization $m=1/(2L)\sum_{i}<s_{i}^{(z)}>$.
From the low frequency behaviour of these quantities one can extract
the spin ($K_{\sigma}$) and charge ($K_{\rho}$) exponents which
characterize the possible presence of gapped phases as: 

\[
K_{\rho}=\frac{\pi}{q}N(q\rightarrow0)\quad,\quad K_{\sigma}=\frac{\pi}{q}S(q\rightarrow0)\quad.\]
\[
\]
We recall that at $m=0$ both in region (ii) and (iii) the ground-state
is a Luther-Emery SC liquid, i.e. it is characterized by $K_{\sigma}=0$
and $K_{\rho}>1$. The actual values of $K_{\rho}$ for $m>0$ , as
obtained from DMRG simulations at two different representative values
of $x$, are given in figures \ref{fig4} and \ref{fig5} (the latter showing
indirectly $K_\sigma$ through the slope at $q=0$).

At low enough magnetization, in both regions we still see evidence
of a closed charge gap, and an open spin gap (not shown). However,
as soon as $m\neq0$ in region (iii) $K_{\rho}<1$ : the SC correlations
are no longer dominant. On the contrary, in region (ii) $K_{\rho}>1$
up to the critical magnetization $m_{1}\approx0.1$ reported in figure
\ref{fig4}. In this case we can conclude that superconductivity is
still present in the magnetized ground state. Accordingly to our previous
discussion, this should imply the presence of FFLO type of oscillations,
which will be investigated in section \ref{secFFLO}. A further interesting
aspect which appears in both regions is the presence of a higher magnetization
value $m_{2}$ above which the system appears to become insulating:
even at finite size, $K_{\rho}$ is zero within the numerical error.
On the other hand our numerical data suggest that within the insulating
regime the spin gap remains open until the magnetization reaches the
value $m_{c}$ at which the breached pair regime is entered. Whereas
for $m>m_{c}$ $K_{\sigma}=1/2$ at variance with the standard Hubbard
case in which $K_{\sigma}=1$ (see figure \ref{fig5}). The result is
a signal of the fully polarized nature of the single electrons for
$m>m_{c}$ in regions (ii) and (iii), which implies that $K_{\sigma}$
coincides with the Luttinger exponent of a gapless spinless liquid,
\textit{i.e.} $1/2$. This is also confirmed by the fact that $S(q)$
manifests a feature at $q=\pi n_{s}$, where $n_{s}$ namely the number
of single electrons, coincides for a fully polarize liquid with $n_{\uparrow}-n_{\downarrow}$.

\begin{figure}
\includegraphics[width=7cm]{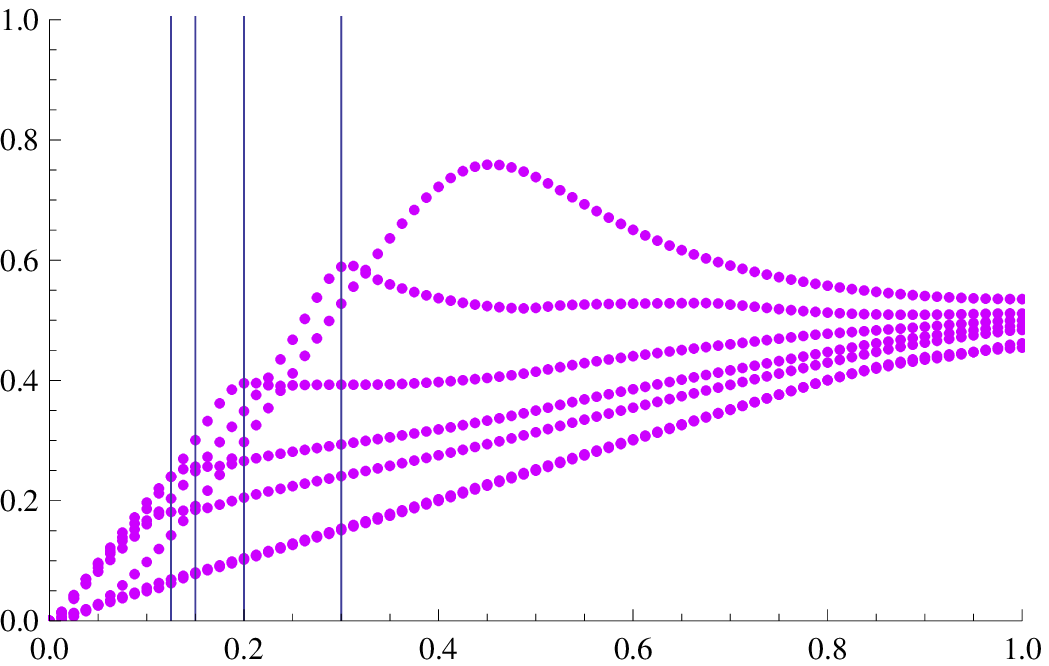}\includegraphics[width=7cm]{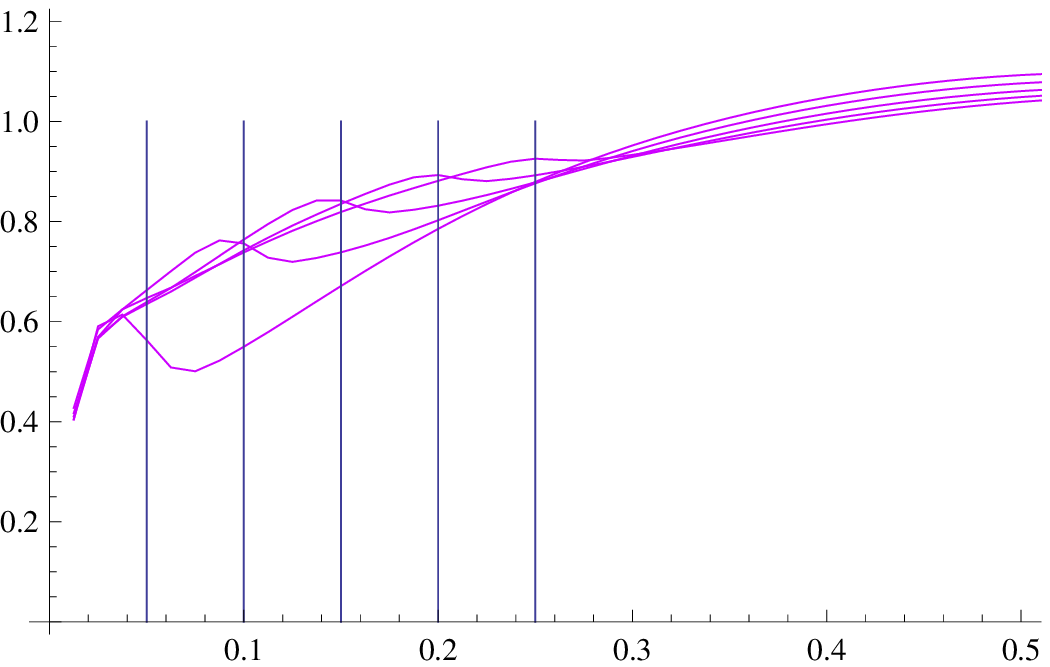}
\caption{Spin structure factor at various values of magnetization $m\leq m_{2}$.
Left panel: $S(q)$ at $x=0.8$, vertical lines in correspondence
of $2n_{s\downarrow}$. Right panel: $S(q)/q$ at $x=0.6$, vertical
lines in correspondence of $2m$ .\label{fig6}}
\end{figure}

To exploit the nature of the difference between region (ii) and (iii)
we also show in figure \ref{fig6} $S(q)$ for $x=0.6,0.8$ at various
magnetization values below $m_{2}$. We recall that in the standard
Luttinger liquid case a feature is expected in $S(q)$ for $q=2\pi n_{\sigma}$\cite{YOA},
whereas a signal of the presence of FFLO would be a feature at $Q=2\pi(n_{\uparrow}-n_{\downarrow})$\cite{RRal}. 

Starting from region (iii), at $x=0.8$ we observe a feature at $q=2\pi n_{s\downarrow}$,
which confirms that in this case the behaviour of magnetic correlations
in the phase is that of a liquid of $n_{s}$ particle, moving in a
background of $n-n_{s}$ empty and doubly occupied sites, in agreement
with our previous observations: in this case, it is just the spin
degrees of freedom of the $n_{s}$ single electrons which rearrange
under the external magnetic field. In region (ii) instead, at $x=0.6$
there is a neat feature at $Q=2\pi(n_{\uparrow}-n_{\downarrow})$,
which is a further signal of the presence of FFLO oscillations. Such
possibility will be explored in section \ref{secFFLO}. The feature
is strongly reduced for $m\gtrsim m_{1}$. In both regions, the feature
disappears at the value $m_{2}$ discussed above.

\section{Phase separation and domain formation}

\begin{figure}
\includegraphics[width=7.2cm]{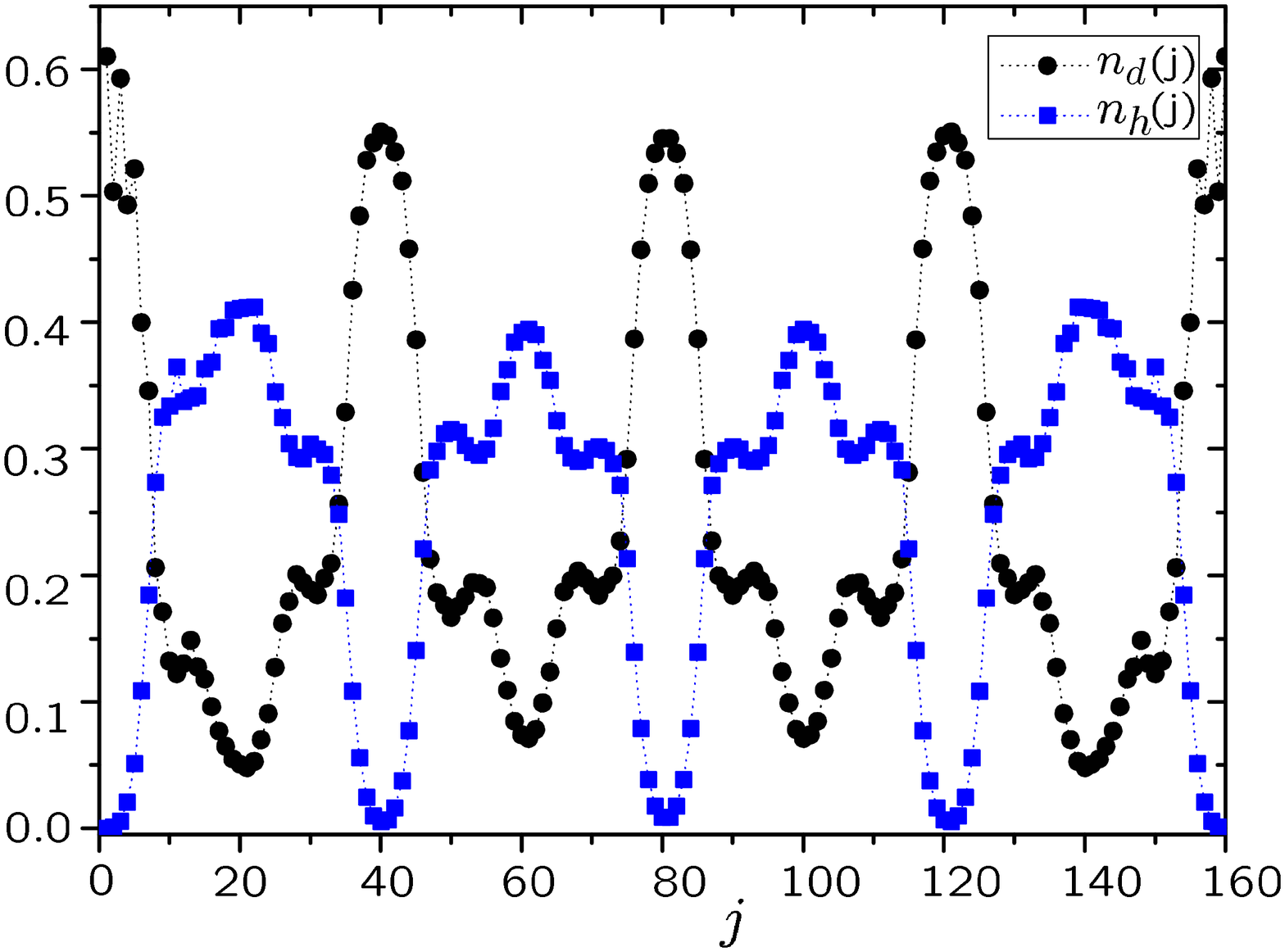}\includegraphics[width=7cm]{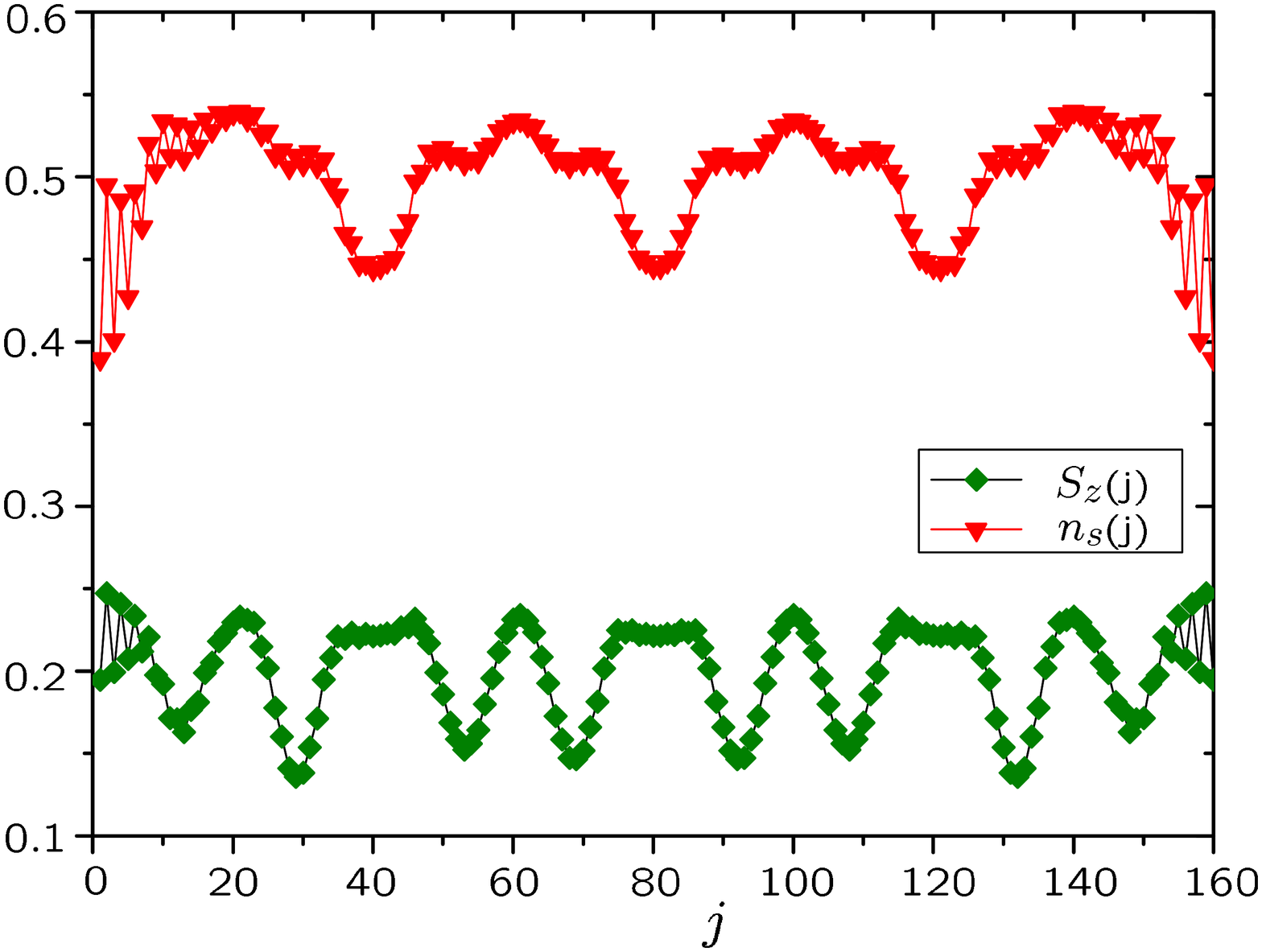}
\caption{Site dependence of various densities for $x=0.8$ (region (iii)) and
$m_{2}\leq m\leq m_{c}$ . Left panel: local densities of holes and
of doublons; right panel: local density of single electrons, and magnetization.\label{fig7}}
\end{figure}

\begin{figure}
\includegraphics[width=7cm]{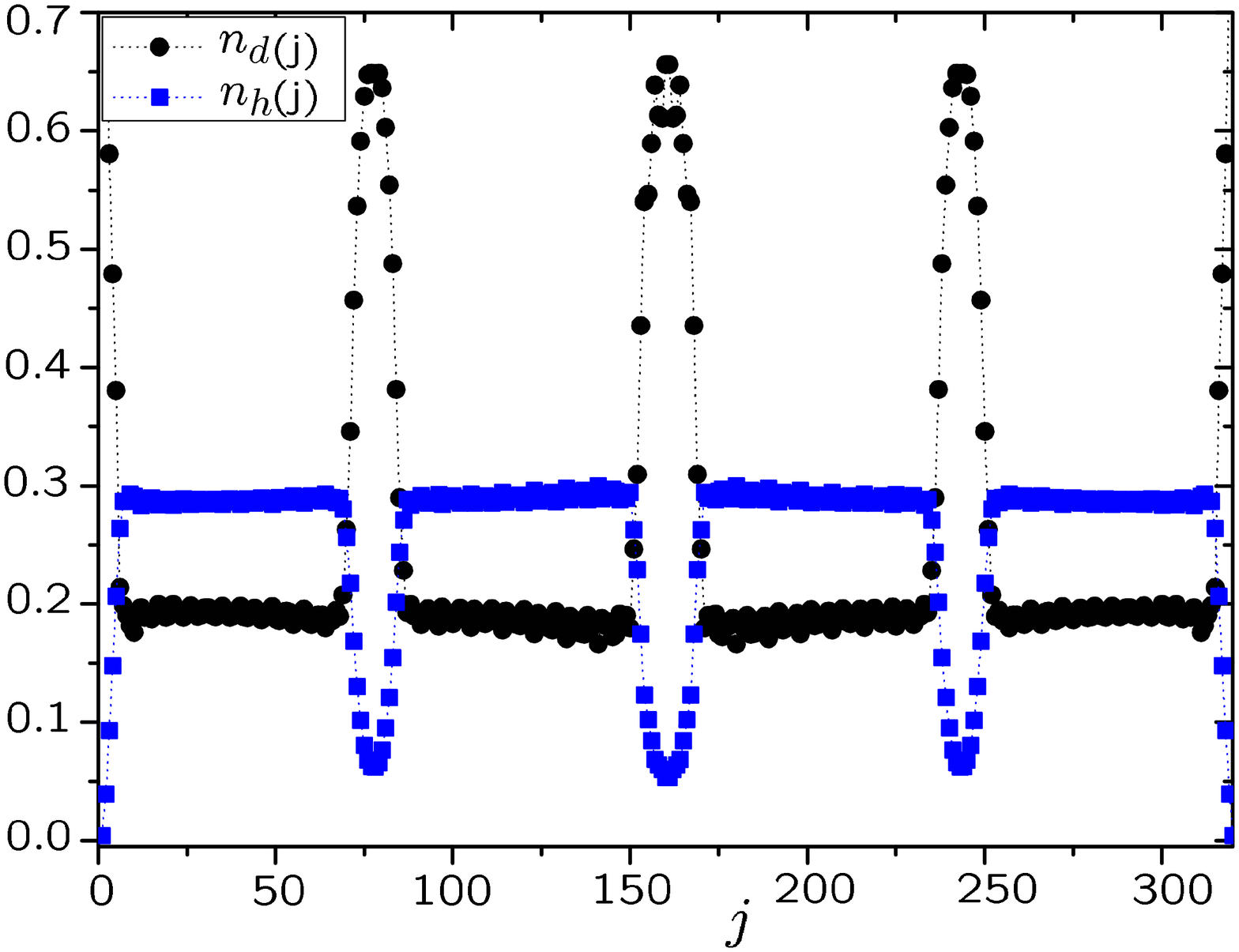}\includegraphics[width=7cm]{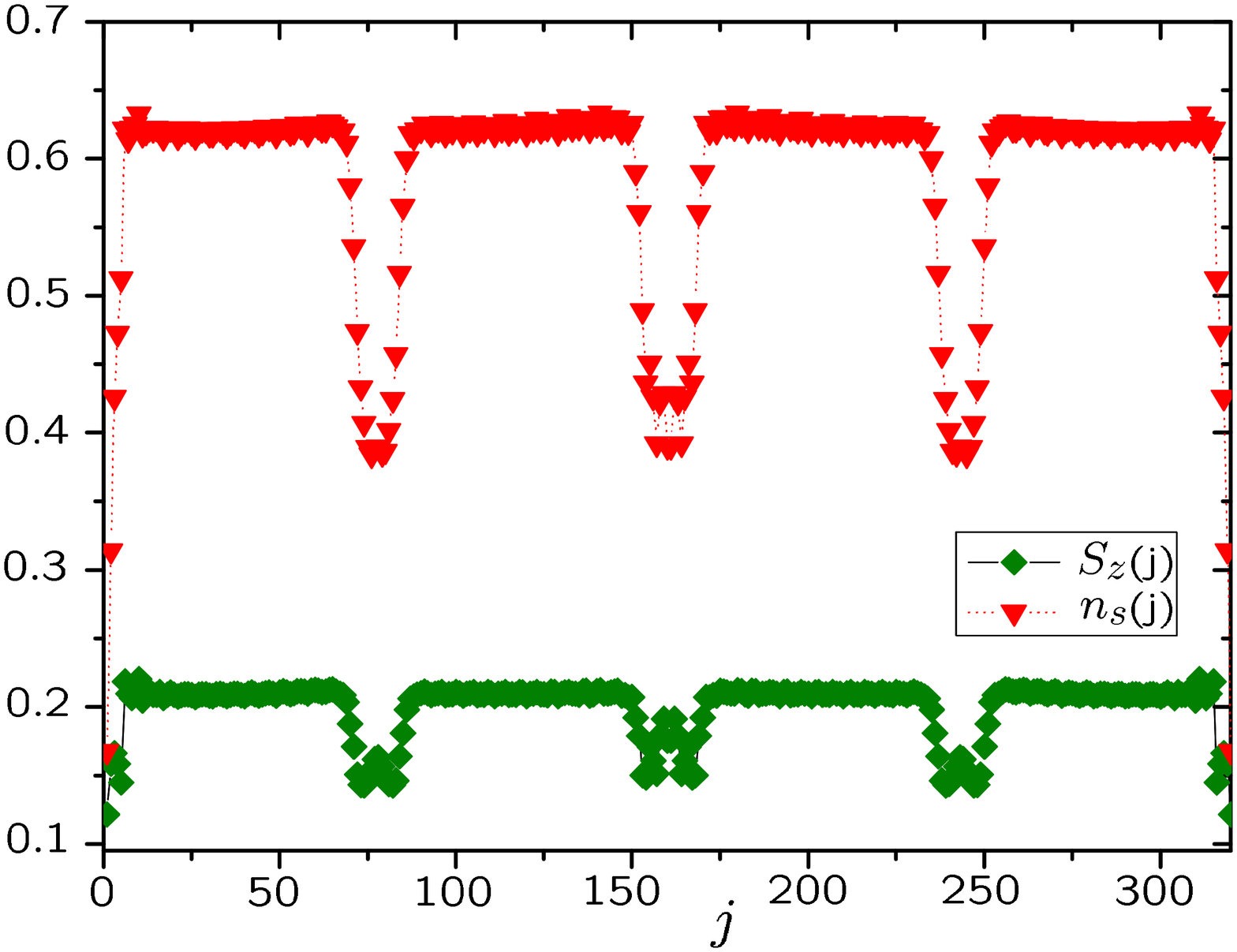}
\caption{Site dependence of local densities of holes and doubly occupied sites
(left panel); single electrons and magnetization (right panel) for
$m_{1}\leq m\leq m_{2}$ at $x=0.6$.\label{fig8}}
\end{figure}

Due to quantum superposition and to the microscopic size of the domains,
at vanishing magnetic field it is not possible to distinguish directly
on the local density $<n(j)>$ and magnetization $<s_{z}(j)>$ profiles
the presence of phase coexistence, even in region (iii) where it can
easily be detected on the chemical potential \cite{ADM}. The problem
persists at low non-vanishing values of the magnetization, in which
case however the local magnetization is observed to display a modulation
with wavelength proportional to either $q$ or $Q,$ depending on
the value of $x$ and consistently with the behaviour observed in $S(q)$
(see previous section). With further increasing the magnetization,
the presence of macroscopic phase separation (MPS) on both the local
density and the local magnetization profiles appears well below the
critical value $m_{c}$ at which the single electrons become fully
polarized. This is seen in figures \ref{fig7} and \ref{fig8}. In fact,
a more careful analysis shows that in region (ii) MPS appears as soon
as $m=m_{1}$ (i.e. $K_{\rho}=1$, where the system looses its SC
properties) . Whereas in region (iii) MPS can be seen only above the
magnetization $m=m_{2}$, in correspondence to the transition to the
insulating state. As also shown from the profiles reported in figure
\ref{fig7} in this case the single electrons of the high density
domains are fully polarized, and simultaneously the holes are expelled
from the same domains, so that the system consists of localized domains
with different average magnetization.

In particular, in figure \ref{fig7} we have chosen to plot separately
in region (iii) the site dependence of the number of doubly $<n_{d}(j)>$
, singly occupied $<n_{s\sigma}^{(i)}>$ , and empty $<n_{h}(j)>$
sites in the different regimes. The presence of low and high-density
domains can be easily recognized: these consists of spatial regions
in which the single electrons move in a background of holes (low density)
or doubly occupied (high density) sites, with different Fermi momenta
(i.e., in this case different values of $n_{s}=n_{s\uparrow}-n_{s\downarrow}$,
shown in figure \ref{fig7}). Since the single electrons can have different
spin orientation only in the low density domains, the site magnetization
turns out to be constant in the high-density ones, whereas it is modulated
according to the number of single electrons with minority spins in
the low density domains. The modulation disappears for $m>m_{c}$,
when the single electrons become fully polarized along all the system.
A consequence of the spatial arrangement is that in the local magnetization
profile the low and high density domains may change their relative
spin orientation depending on the value of $m$. Indeed, as soon as $m>m_{2}$
the system chooses to lower the magnetization of the low density domains,
which weight is more relevant in the kinetic energy. Increasing $m$
then amounts to align more and more spins of the electron in the low
density domains, which at $m=m_{2}$ are more numerous than those
in the high density domains, so that the relative magnetization is
reversed. Recalling that the high and low density domains are characterized
respectively by the density of single electrons $n_{sh}$, and $n_{sl}$,
in this case it is possible to calculate explicitly $m_{c}$: $m_{c}=n_{sl}-(n_{sh}-1)$.

As mentioned, in region (ii) it is possible to distinguish the presence
of high and low density domains already at lower values of magnetization
(figure \ref{fig8}), as soon as SC correlations cease to be dominant
in the whole system ($m=m_{1}$). In this case the single electrons
with opposite spin orientation vary coherently with the sites in the
low density domains, so that their difference (hence the local magnetization)
remains constant there. We will see later that this is a further signal
of the presence of FFLO in these domains. Whereas the local magnetization
appears to be modulated in the high density domains, since the single
electrons are not yet fully polarized. The low density phase is different
from the one in region (iii), and the coherent behaviour of electrons
with opposite spin suggest that in correspondence to the appearance
of MPS SC pairs become confined to these domains. Further increasing
$m$ above $m_{2}$ the behaviour resembles that of the $x=0.8$ case:
in both regions the system behaves \textit{de facto} as an insulating
ferrimagnet.

Finally for $m>m_{c}$ the breached pairs regime is entered. Again,
this fate is shared also by region (ii).

\section{Pair-pair correlations and FFLO oscillations}\label{secFFLO}

In this section we consider the possible presence of FFLO oscillation
in pair pair correlations at non-zero magnetization via the study
of various pairing correlators $\langle A_{i}^{\dagger}A_{i+R}\rangle$
where the operator $A_{i}$ at site may refer to $A_{i}^{{\rm onsite}}=\eta_{i}=c_{i\downarrow}c_{i\uparrow}$
(onsite pairing) or to singlet and triplet combinations on adjacent
sites $i$ and $i+1$\[
A_{i}^{{\rm sing}}=S_{i}=\frac{1}{\sqrt{2}}(c_{i+1\downarrow}c_{i\uparrow}-c_{i+i\uparrow}c_{i\downarrow})\]
 \[
A_{i}^{{\rm trip-0}}=T_{0i}=\frac{1}{\sqrt{2}}(c_{i+1\downarrow}c_{i\uparrow}+c_{i+i\uparrow}c_{i\downarrow})\]

\begin{equation}
A_{i}^{{\rm trip}-\sigma}=T_{\pm1i}=c_{i+1\sigma}c_{i\sigma}\;,\;\;\sigma=\uparrow,\downarrow\label{eq:AAA}
\end{equation}

For each one of these operators, as well as for their correlators,
we can use Hubbard operators $X$ and give a decomposition based on
the high-density subspace made only of doubly occupied sites and singly
occupied sites ($2\sigma$ elements), the low-density subspace where
no doubly occupied sites exist while empty sites are permitted ($0\sigma$
elements). Mixed contributions also must be considered, as can be
seen from the example of $T_{1i}=T_{1i}^{{\rm low}}+T_{1i}^{{\rm high}}+T_{1i}^{{\rm mixed}}$
with\[
T_{1i}^{{\rm low}}=X_{i+1}^{0\uparrow}X_{i}^{0\uparrow}\;,\;\; T_{1i}^{{\rm high}}=X_{i+1}^{\downarrow2}X_{i}^{\downarrow2}\]
 \[
T_{1}^{{\rm mixed}}=X_{i+1}^{\downarrow2}X_{i}^{0\uparrow}+X_{i+1}^{0\uparrow}X_{i}^{\downarrow2}\]
For $T_{-1i}$ it is sufficient to reverse the index $\sigma$ of
single occupation. The 0 component of the triplet and the singlet
have similar, although longer, expressions that we do not report here
for the sake of brevity.

\begin{figure}
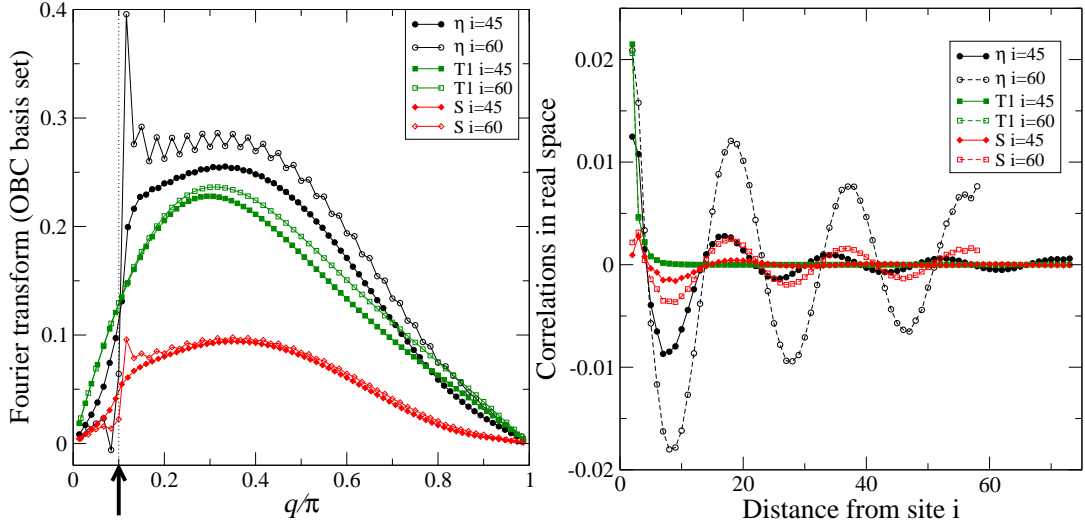

\includegraphics[width=7cm,clip]{fig9l}\includegraphics[width=7.3cm,clip]{fig9r}
\caption{Various types of correlators $\langle A_{i}^{\dagger}A_{i+R}\rangle$
with the choices for $A_{i}$ reported in the legend (compare with
(\ref{eq:AAA})). Considering the emerging PS we have selected
two possible starting sites $i=45$ or $i=60$. In the left hand side
the correlation functions are plotted in reciprocal space and the
arrow marks the expected FFLO peak at $q_{{\rm FFLO}}=k_{F\uparrow}-k_{F\downarrow}=\pi(n_{\uparrow}-n_{\downarrow})=2\pi m_{z}$.
Here the magnetization density, that allows for the FFLO effect, is
$m_{z}=0.05$. The oscillating pattern in real space is evident from
the right panel, especially for the dominating $\eta$-pairs correlation
that decay slowly. The $T_{1}$ correlations instead do not exhibit
the same effect. The model parameters here are $u=0$ and $x=0.6$.\label{fig:FFLOx0.6}}
\end{figure}

\begin{figure}
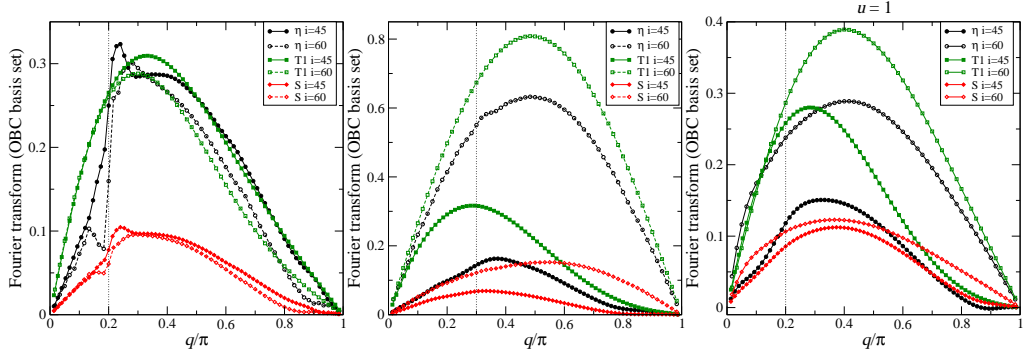

\includegraphics[width=4.5cm,clip]{fig10l}\includegraphics[width=4.5cm,clip]{fig10c}\includegraphics[width=4.5cm,clip]{fig10r}
\caption{Left and central panel: as figure \ref{fig:FFLOx0.6} but with magnetization
density increased to $0.1$ and $0.15$ respectively. In the former
case a weak signature of FFLO peak remains (it would be expected in
correspondence of the dotted vertical line), while in the latter case
is wiped out. The same happens by increasing the Coulomb interaction
at the same value of $m_{z}=0.1$ (for instance $u=1$ in the example
of the right panel).\label{fig:FFLOwipeout}}
\end{figure}

We discuss the results of a quantitative DMRG analysis by choosing
$x=0.6$ and $x=0.8$ as representative parameters of the two somehow
different situations of region (ii) and (iii) respectively. All the
numerical data presented in this section are taken on chains of $L=120$
with open boundary conditions (OBC) sites at half filling. First,
in figure \ref{fig:FFLOx0.6} we show evidence of FFLO behaviour showing
up when the populations are slightly imbalanced in region (ii). In
figure \ref{fig:FFLOwipeout}, instead, we can see how the FFLO effect
is progressively wiped out if either the magnetization density is
too high (central panel), or if the interaction $u$ is increased
(right panel).

\begin{figure}
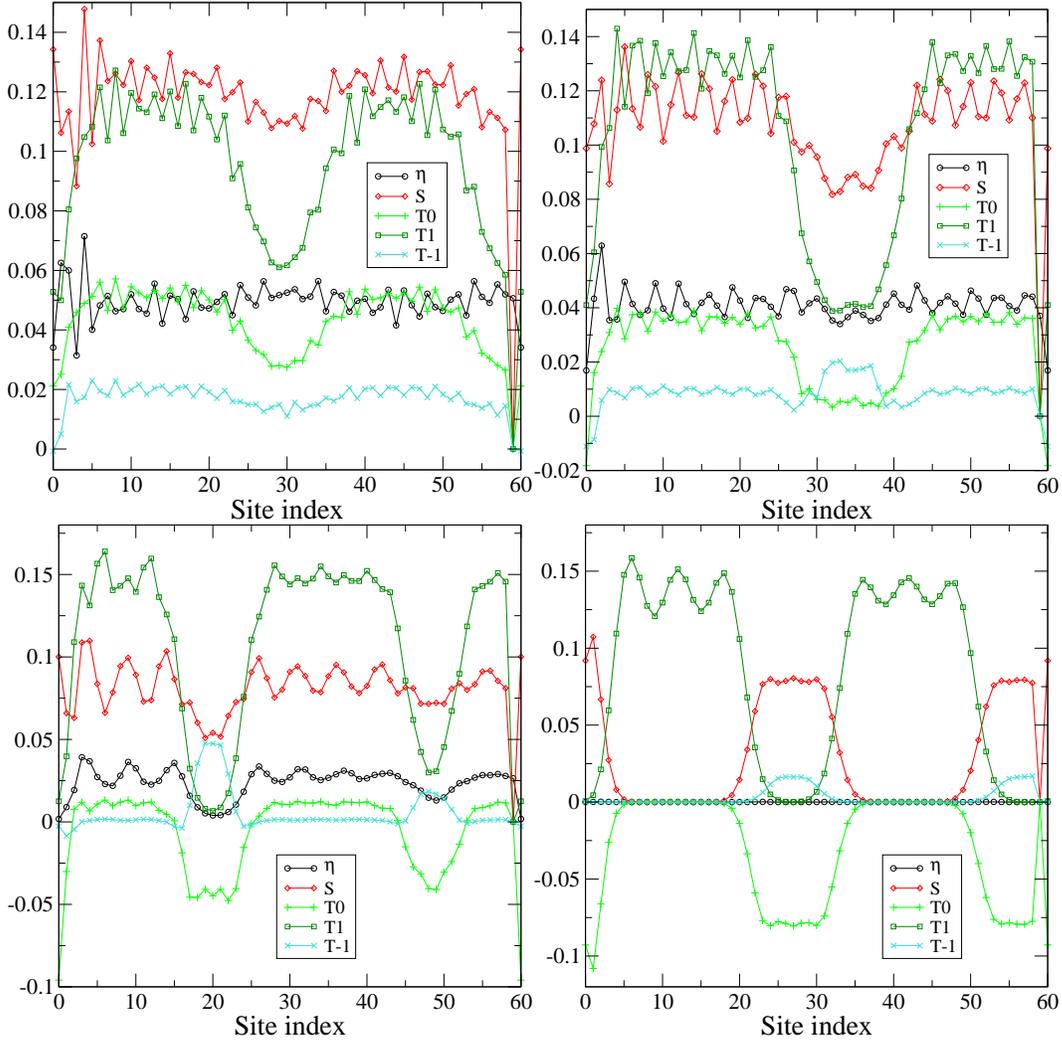

\includegraphics[width=7cm,clip]{fig11ul}\includegraphics[width=7cm,clip]{fig11ur}

\includegraphics[width=7cm,clip]{fig11dl}\includegraphics[width=7cm,clip]{fig11dr}
\caption{Various types of correlators $\langle A_{i}^{\dagger}A_{i+R}\rangle$
with the choices for $A_{i}$ written in (\ref{eq:AAA}) and the
distance fixed to $R=1$. The model parameters here are $u=1$, $x=0.6$
and with OBC one can appreciate the spatial dependence on $i$. Upper
panel: $m_{z}=0.1$ (left) and $m_{z}=0.15$ (right); lower panel:
$m_{z}=0.25$ (left) and $m_{z}=0.40$ (right). Due to reflection
symmetry only the left half of the chain is displayed.\label{fig:R1}}

\end{figure}

Due to the possible emergence of PS, that is spatial inhomogeneities,
it will be useful to inspect also the variation of the correlators
$\langle A_{i}^{\dagger}A_{i+R}\rangle$ along the chain at fixed
$R$ as a function of $i$. When $R=1$ there is a site $i+1$ in
common between $A_{i}^{\dagger}$ and $A_{i+1}$ (apart from the case
$A_{i}=\eta_{i}$) and there are some simplifications in the following
correlation functions referred to the {}``high'' and {}``low''
sectors of the Hilbert space\[
\langle T_{\pm i}^{{\rm low}\dagger}T_{\pm i+1}^{{\rm low}}\rangle=-\langle X_{i}^{\sigma0}n_{s\sigma i+1}X_{i+2}^{0\uparrow}\rangle\]
 \[
\langle T_{\pm i}^{{\rm high}\dagger}T_{\pm i+1}^{{\rm high}}\rangle=-\langle X_{i}^{2\bar{\sigma}}n_{di+1}X_{i+2}^{\bar{\sigma}2}\rangle\]
 where $\pm$ in the subscript corresponds to $\sigma=\uparrow,\downarrow$
and $\bar{\sigma}$ means spin flip. In addition, it turns out that
the {}``high'' parts of the singlet and of the 0-component triplet
are always equal except for the sign\[
\langle S_{i}^{{\rm high}\dagger}S_{i+1}^{{\rm high}}\rangle=\frac{1}{2}(\langle X_{i}^{2\downarrow}n_{di+1}X_{i+2}^{\downarrow2}\rangle+\langle X_{i}^{2\uparrow}n_{di+1}X_{i+2}^{\uparrow2}\rangle)\]
 \begin{equation}
\langle T_{0i}^{{\rm high}\dagger}T_{0i+1}^{{\rm high}}\rangle=-\langle S_{i}^{{\rm high}\dagger}S_{i+1}^{{\rm high}}\rangle\label{eq:ST0h}\end{equation}
In figure \ref{fig:R1} we examine the local dependence of pairing correlations
along a chain showing how PS appears also in SC properties. The relative
distance is now fixed to $R=1$ site and the leftmost site $i$ is
varied; again we select the example $u=1$, $x=0.6$ as in figure \ref{fig:FFLOwipeout}
(right). As long as the net magnetization density is small the dominant
pairing correlations are of singlet type and the dependence on $i$
is not strong. When the magnetization is increased the $+1$ component
of the triplet starts to dominate, but in the high-density regions
the pairing correlations are generally suppressed, with the exception
of a small enhancement of the $-1$ component of the triplet. Notice
that we may arrive at a situation in which there are no unpaired down
particles (bottom right panel) and, correspondingly, the singlet and
$T_{0}$ correlators acquire the same profile except for the sign
as in (\ref{eq:ST0h}). 

\begin{figure}
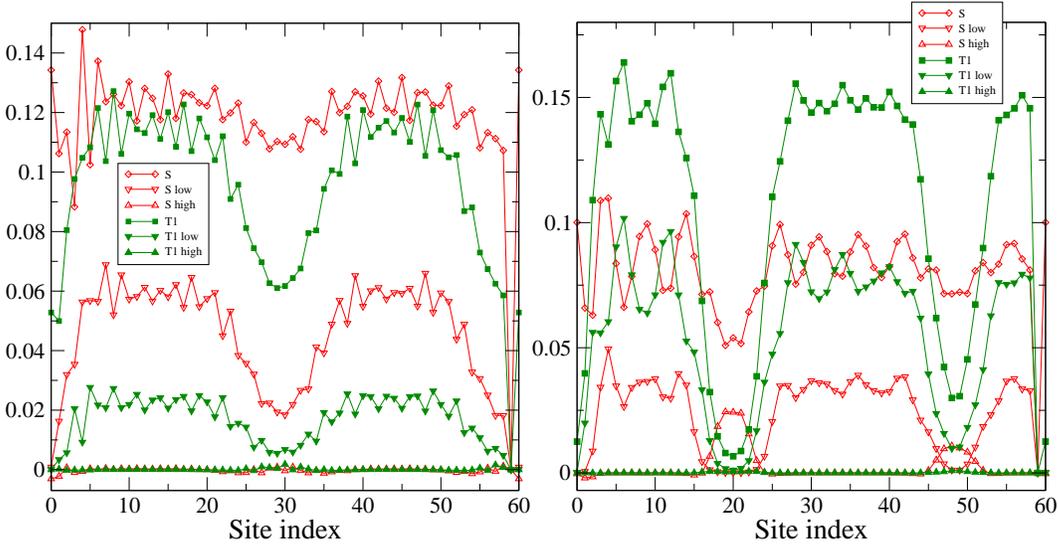

\includegraphics[width=7cm,clip]{fig12l}\includegraphics[width=7cm,clip]{fig12r}
\caption{Same as figure \ref{fig:R1} for the singlet and $T_{+1}$ correlation
functions at distance $R=1$, where full up triangles now denote the
{}``high'' component and the empty down triangles denote the {}``low''
one. Left panel is for $m_{z}=0.1$ while the right one is for $m_{z}=0.25$.\label{fig:R1lh}}

\end{figure}

Having identified the dominant type of pairing correlations, at least
at short distance, we can also inspect the separate contributions
from low- and high-density sectors of the Hilbert space to the singlet
and $T_{+1}$ correlators, according to expressions above. This is
done in figure \ref{fig:R1lh}.

Qualitatively the result can be summarized as follows. On increasing
the magnetization $m$ the spin-up electrons align along the field
first in the high-density islands and then in the low-density ones.
When in region (ii) the system exhibits FFLO oscillations in pair
structure factor. When the domains finally appear at a macroscopic
level ($m=m_{1}$), the FFLO pairs become mostly localized into the
low density domains, and the corresponding peak is weakened; globally
the system behaves as metal made of ferromagnetic domains alternating
with superconducting ones. The SC properties of the low density domains
fully disappear only when single electron localize ($m>m_{2}$), as
shown by the right panel of figure \ref{fig:R1}.

\section{Discussion and Conclusions}

\begin{figure}
\includegraphics[width=9cm]{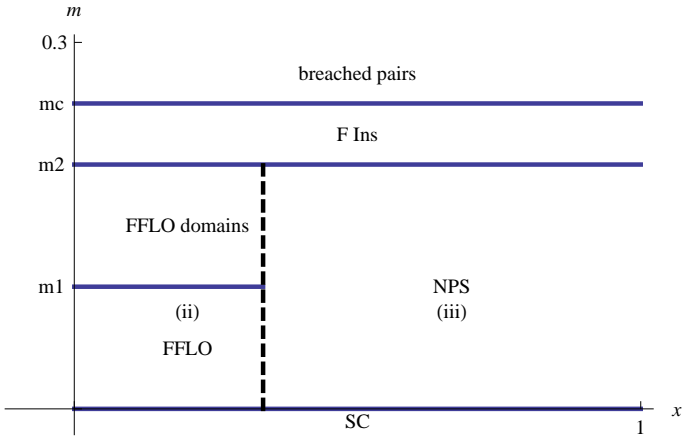}
\caption{Schematic phase diagram at $0\le u<u_{c}(x)$ and half-filling, in
the $x-m$ plane for $x\geq1/2$}

\label{fig13}
\end{figure}

The results of the previous sections provide a coherent scenario of
the physics of the SC phase described by the Hamiltonian (\ref{eq:1})
for $x\gtrsim1/2$ in presence of non-vanishing magnetization. With
respect to the case $x\leq1/2$ (region (i)) , the different behaviour
observed previously at $m=0$ persists up to high magnetization values
($m>m_{c}$), as a consequence of the rearrangement in the kinetic
energies of holes and doublons which takes place at $x=0.5$ (see
section II). The main features of the phase diagram in the $x-m$
plane are schematically shown in figure \ref{fig13} for regions (ii)
and (iii).

At low magnetization the diagram shows evidence of the two regions,
which turn out to behave differently up to $m=m_{2}$; here $m_{2}$
is the value at which the holes and minority spin single electrons
localize into the low density domains, and the doublons localize into
the high density domains: the system becomes insulating. In region
(ii) SC persists in the whole system at non-zero magnetization up
to $m=m_{1}<m_{2}$; the pair structure factor correspondingly shows
a strong FFLO peak. Macroscopic phase separation appears on both the
density and the magnetization profiles for $m>m_{1}$: in this case
SC correlations become mainly confined to the low-density domains,
whereas the system is still metallic due to the mobility of the single
electrons. The low and high density regions display different magnetizations,
so that for $m_{1}\leq m\leq m_{2}$ the system can be considered
an itinerant ferromagnet, with superconducting FFLO domains. As for
region (iii), there is no evidence for dominant SC correlations in
the NPS phase as soon as $m\neq0$; the system behaves as a liquid
of $n_{s}$ single particles moving in a background of empty and doubly
occupied sites textured into domains of microscopic size\cite{ADM}.
Macroscopic phase separation into high- and low-density domains appears
on the density and magnetization profiles only for $m>m_{2}$, simultaneously
with the transition to the insulating state. Finally, at a higher
magnetization $m_{c}$, the regime of breached pairs described in
\cite{ABM} is entered.

We emphasize that the presence of a FFLO regime in the repulsive Hubbard
model takes place for a range of values of the off-diagonal Coulomb
repulsion ($1/2\lesssim x\lesssim2/3$) in principle accessible to
experiments. The results reported here for such regime are complementary
to those obtained in \cite{HKYT} for the extended Hubbard model; in fact, they are expected to persist also in presence of non-vanishing nearest neighbor Coulomb repulsion, since similar incommensurate spin and charge correlations are observed also in that case \cite{AGA}.  
The results are reminiscent for some aspects of those reported in \cite{LoTr}
for the standard Hubbard model in the attractive $u$ region. We guess
that our FFLO regime is to be identified to the weak LO (Larkin-Ovchinnikov)
regime described there, whereas the regime of macroscopic phase separation
of the high density polarized walls and the low density FFLO domains
could well coincide with the strong LO regime discussed there, in
which pair-pair correlations exhibit spatial dependence. The main
difference we see with \cite{LoTr}is that in the present case with
increasing the magnetization the domain walls (to be identified with
our high density domains) tend to localize, in contrast to the delocalized
behaviour observed there. Ultimately this feature is responsible for
the transition to the insulating regime found here. We expect that
the two scenarios could merge in the region of weakly attractive $u$
and $x\leq0.5$.

As mentioned, with varying magnetization a second transition then
occurs in both regions at $m=m_{2}$; in this case the system consists
of two segregated metals of fully polarized single electrons with
different Fermi momenta, which as a whole behave as a (ferrimagnetic)
insulator. In two dimensions, we argue that the behaviour of such insulator
could well be that of metallic stripes; whereas for $m_{1}\leq m\leq m_{2}$
in region (ii) the physics chould be that of a striped superconductor
\cite{BFKT}. 

The analysis is performed here at finite size. Due to the different
qualitative behaviour of the phases depicted in figure \ref{fig13} the
scenario should be considered quite plausible. A finite size scaling
analysis is expected to provide a quantitative accurate derivation
of the critical lines. In particular, since our study just considered
two far apart typical points ($x=0.6$ for region (ii), and $x=0.8\mbox{ }$
for region (iii)), it is not possible to infer from it whether the
change from region (ii) to (iii) is a smooth crossover or a true transition
line.
\ack
We thank the Bologna Section of the INFN for the computational resources.
AM acknowledges stimulating discussions with D.C. Campbell, C. Chamon,
and N. Trivedi, as well as the hospitality of Condensed Matter Theory's
Visitor Program at Boston University, where this work was completed.
We are also grateful to F. Ortolani for the DMRG code. The work was
partially supported by national italian funds, PRIN2007JHLPEZ 005.

\end{document}